\documentclass{article}
\usepackage{graphicx}
\usepackage{amsmath}
\usepackage[round]{natbib}
\usepackage{authblk}
\title{Dissipative models of swell propagation across the Pacific}
\author[1]{Camille R. Zaug}
\author[1*]{John D. Carter}
\affil[1]{Mathematics Department; Seattle University; Seattle, WA; USA}
\affil[*]{Corresponding author, carterj1@seattleu.edu}

\begin{document}

\maketitle

\begin{abstract}
    Ocean swell plays an important role in the transport of energy across the ocean, yet its evolution is not well understood.  In the late 1960s, the nonlinear Schr{\"o}dinger (NLS) equation was derived as a simplified model for the propagation of  ocean swell over large distances.  More recently, a number of dissipative generalizations of the NLS equation based on a simple dissipation assumption have been proposed.  These models have been shown to accurately model wave evolution in the laboratory setting, but their validity in modeling ocean swell has not previously been examined.  We study the efficacy of the NLS equation and four of its generalizations in modeling the evolution of swell in the ocean.  The dissipative models perform significantly better than conservative ones and are overall reasonable models for swell amplitudes, indicating dissipation is an important physical effect in ocean swell evolution.  The nonlinear models did not out-perform their linearizations, indicating linear models may be sufficient in modeling some aspects of ocean swell evolution over large distances.
\end{abstract}

\section{Introduction}

Swell in the ocean is composed of slowly modulated surface wave trains with relatively long periods.  It is typically formed after waves created by distant storms have had a chance to disperse.  Swell can travel thousands of kilometers, see for example, \cite{Snodgrass} and \cite{Collard}.  This coherence over long distances might suggest that there may be a simple underlying model that governs the evolution of swell.  However, \cite{Rogers} and \cite{Rascle} show that swell amplitudes are relatively poorly predicted.  Although \cite{Snodgrass} neglected dissipative effects, more recent work suggests that dissipation may play an important role in swell evolution, see for example, \cite{Collard}, \cite{Ardhuin}, \cite{HendersonSegur}, and \cite{YoungBabanin}.  

\subsection{Model Equations}

The one-dimensional cubic nonlinear Schr\"odinger (NLS) equation,
\begin{equation}
    iu_{\chi}+u_{\xi \xi}+4|u|^2u=0,
    \label{NLS}
\end{equation}
is an approximate model for the slow evolution of a nearly monochromatic wave train of gravity waves on deep water in one dimension (i.e.~swell propagating in one direction over large distances).  Here $u=u(\xi,\chi)$ is a dimensionless complex-valued function that describes the evolution of the envelope of the oscillations of a carrier wave, $\chi$ represents dimensionless distance across the ocean, and $\xi$ represents dimensionless time.  The leading-order approximation to the dimensional surface displacement, $\eta(x,t)$, can be obtained from an NLS solution, $u(\xi,\chi)$, via the relation
\begin{equation}
    \eta(x,t)=\frac{\epsilon}{k_0}\left( u(\xi,\chi)\mbox{e}^{i\omega_0t-ik_0x}+u^*(\xi,\chi)\mbox{e}^{-i\omega_0t+ik_0x}\right)+\mathcal{O}(\epsilon^2),
    \label{SurfaceDisplacement}
\end{equation}
where $\eta$, $x$, and $t$ are dimensional variables and $u^*$ represents complex conjugate of $u$.  Here $\omega_0$, $k_0$, and $a_0$ are parameters that represent the dimensional frequency, wavenumber, and amplitude of the carrier wave respectively, and $\epsilon=2a_0k_0$ is a dimensionless measure of wave steepness/nonlinearity.  The dimensionless and dimensional independent variables are related by
\begin{equation}
    \xi=\epsilon\omega_0 t-2\epsilon k_0x,\hspace*{1cm}\chi=\epsilon^2k_0x.
\end{equation}

\cite{Zak1968} derived the NLS equation as a model for the propagation of ocean swell over large distances.  See \cite{Johnson} for a more detailed and modern derivation, but  note that both of these derivations rely on an ansatz that is slightly different than the one given in equation (\ref{SurfaceDisplacement}).  In deriving the NLS equation and the generalizations presented below, one assumes that the surface displacement is small, that the spectrum is narrow banded, and that the spectrum is centered about a dominant frequency known as the carrier wave.  The NLS equation has been studied extensively from a mathematical perspective, see for example, \cite{AS} and \cite{SulemSulem}.  The NLS equation has also been shown to favorably predict measurements from laboratory experiments when the waves have small amplitude and steepness (i.e.~$\epsilon<0.1$), see for example, \cite{LoMei}.

In order to weaken the NLS equation's narrow-bandedness restriction, \cite{Dysthe} extended the NLS derivation asymptotics one additional order and derived the equation that now bears his name
\begin{equation}
    iu_{\chi}+u_{\xi \xi}+4|u|^2u+\epsilon \big(-8iu^2u^*_{\xi}-32i|u|^2u_{\xi}-8iu\big(\mathcal{H}(|u|^2)\big)_{\xi}\big)=0.
    \label{Dysthe}
\end{equation}
Here $\mathcal{H}$ represents the Hilbert transform, which is defined by
\begin{equation}
    \mathcal{H}\left(f(\xi)\right) = \sum_{k=-\infty}^{\infty} -i \text{sgn}(k)\hat{f}(k)e^{2 \pi i k \xi/L},
    \label{Hilbert}
\end{equation}
where $\hat{f}(k)$ is the Fourier transform of the function $f(\xi)$ and is defined by
\begin{equation}
    \hat{f}(k)=\frac{1}{L}\int_0^L f(\xi)\mbox{e}^{-2\pi ik\xi/L}d\xi,
    \label{Fourier}
\end{equation}
and $L$ is the $\xi$-period of the measurements.  \cite{LoMei} showed that the Dysthe equation accurately predicts laboratory experimental measurements for a wider range of wave amplitude and steepness values than does the NLS equation.

Neither the NLS equation nor the Dysthe equation include terms that account for decay and/or dissipation.  In other words, both are conservative partial differential equations (PDEs).  In order to address this limitation, a number of dissipative generalizations of the NLS equation have been proposed and studied.  In this work, we focus on three dissipative generalizations of the NLS equation.  \cite{StabBF} and \cite{Wu} showed that predictions obtained from the dissipative nonlinear Schr\"odinger (dNLS) equation
\begin{equation}
    iu_{\chi}+u_{\xi \xi}+4|u|^2u+i\delta u=0,
    \label{dNLS}
\end{equation} 
where $\delta$ is a nonnegative constant representing {\emph{all}} dissipation and decay, compared favorably with a range of laboratory experiments.  In this model and those included below, dissipation and decay from {\emph{all}} sources (including viscosity, wave breaking, geometric spreading, air-sea stresses, etc.) are accounted for by the single parameter, $\delta$.  This is the simplest dissipative generalization of the NLS equation as the dissipation is constant and frequency independent.  While \cite{YoungBabanin} showed that the ocean swell decay rate is proportional to the wavenumber squared, it is a reasonable first-order assumption that the dissipation rate is wavenumber independent since the ocean data we examine is narrow banded and the models rely on a narrow-bandedness assumption.  \cite{HendersonSegur} used the dNLS equation as a basis for a comparison of dissipation rates, frequency downshift, and evolution of swell in laboratory experiments and in the ocean using the \cite{Snodgrass} data.

Recently, following the work of \cite{Dysthe} and \cite{DDZ}, \cite{FD1} derived the viscous Dysthe (vDysthe) equation
\begin{equation}
    iu_{\chi}+u_{\xi \xi}+4|u|^2u+i\delta u+\epsilon \big(-8iu^2u^*_{\xi}-32i|u|^2u_{\xi}-8iu\big(\mathcal{H}(|u|^2)\big)_{\xi}+5\delta u_\xi\big)=0,
    \label{vDysthe}
\end{equation}
from the dissipative generalization of the water-wave problem presented by \cite{Wu}.  Additionally, they showed that the vDysthe equation accurately predicts the evolution of slowly-modulated wave trains from two series of experiments.  We note that much of the \cite{DDZ} work had been done previously independently by \cite{Lundgren}.  In this model, the dissipation rate depends linearly on the wavenumber.  A flaw arises in the vDysthe equation because of this linear dissipation rate: The amplitude of any lower sideband with dimensionless frequency further than $1/(5\epsilon)$ from the carrier wave will grow exponentially in $\chi$.  This is obviously not physical.  The flaw results from limitations associated with the narrow-bandwidth assumption used in the derivation of the vDysthe equation.  See Section 2.2 of \cite{FD2} for more details.

Motivated by the work of \cite{GramstadTrulsen}, \cite{FD2} showed that the ad-hoc dissipative Gramstad-Trulsen (dGT) equation,
\begin{equation}
    iu_{\chi}+u_{\xi \xi}+4|u|^2u+i\delta u+ \epsilon\big(-32i|u|^2u_\xi-8u\big(\mathcal{H}(|u|^2)\big)_{\xi}+5\delta u_\xi\big)- 10i\epsilon^2\delta u_{\xi\xi}=0,
    \label{dGT}
\end{equation}
accurately predicts the evolution of slowly-modulated wave trains from four series of laboratory experiments.  In this model, the dissipation rate depends quadratically on the wavenumber,  which is at least qualitatively similar with the observations of \cite{YoungBabanin}.  Although the accuracy of the vDysthe and dGT equations were similar for the experiments examined, it is important to note that the dGT equation does not have the same nonphysical growth flaw as the vDysthe equation because of the addition of the $\epsilon^2\delta u_{\xi\xi}$ term.   

{\emph{The main goal of this paper is to test the accuracy of the NLS, Dysthe, dNLS, viscous Dysthe, and dissipative Gramstad-Trulsen equations as one-dimensional models of swell traveling across the Pacific Ocean.}}

\subsection{Frequency Downshift}

Frequency downshift (FD) is said to occur when the carrier wave loses a significant amount of energy to its lower sidebands.  FD was first observed in wave-tank experiments conducted by \cite{LakeEtAl} and \cite{LakeYuen}.  Using a wave maker located at one end of the tank, they created a wave train with a particular frequency.  As the waves traveled down the tank, they experienced the growth of the \cite{BF} instability and disintegrated.  Further down the tank, the waves regained coherence and coalesced into a wave train with a lower frequency than the one created by the wave maker.

There are two common metrics used to quantify FD: a monotonic decrease in the wave's spectral peak or a monotonic decrease in the wave's spectral mean.  FD is said to be temporary if an initial decrease in either the spectral peak or mean is followed by an increase.  The spectral peak, $\omega_p$, is defined to be the frequency with maximal amplitude.  The spectral mean, $\omega_m$, is defined by
\begin{equation}
    \omega_m = \frac{\mathcal{P}}{\mathcal{M}},
    \label{omegam}
\end{equation}
where $\mathcal{P}$ describes the ``linear momentum'' of the wave and is given by
\begin{equation}
    \mathcal{P} =\frac{i}{2L}\int_0^L(uu_\xi^*-u_\xi u^*)d\xi,
    \label{Pnorm}
\end{equation}
and $\mathcal{M}$ describes the ``mass'' of the wave and is given by 
\begin{equation}
    \mathcal{M} = \frac{1}{L} \int_0^L|u|^2d\xi,
    \label{Mnorm}
\end{equation}
where $L$ is the period of the $\xi$ measurement.  Since $\omega_p$ is a ``local'' frequency measurement and $\omega_m$ is a ``global'' frequency measurement, it is possible for a particular wave train to exhibit FD in neither, either, or both senses.  The experiments of \cite{LakeEtAl} and \cite{LakeYuen} provide a clear demonstration of FD in the spectral peak sense.  Their experiments also likely exhibited FD in the spectral mean sense, but neither $\mathcal{P}$ nor $\omega_m$ was measured, so a definitive statement regarding FD in the spectral mean sense for those experiments cannot be made.  Most physical explanations for FD rely on wind and wave breaking, see for example \cite{TD}, \cite{HaraMei}, and \cite{BrunettiEtAl}.  However, the laboratory experiments examined by \cite{StabBF} exhibited FD in both senses without wind or wave breaking.  Thus, there must be a mechanism for this phenomenon which does not rely on these effects.  The dGT and vDysthe equations, which predict FD in the spectral mean sense without relying on wind or wave breaking, were proposed as models for this phenomenon.

The ocean swell data collected on the Pacific Ocean by \cite{Snodgrass} displays evidence of FD in both senses.  We examine this data in detail in Section \ref{OceanData}.  Understanding the mechanisms for FD will contribute to knowledge about how energy propagates across the ocean, with the potential to improve predictive abilities as swell nears the shore, impacting fields from shipping to surfing.  {\emph{A secondary goal of this paper is to examine frequency downshift in ocean swell data and in these models.}}

\subsection{Model Properties}

There is no mathematical theory that governs the evolution of the spectral peak over large distances for any of the PDEs under consideration.  However, the theory for the evolution of the spectral mean for these equations is well known.  The NLS equation preserves both $\mathcal{M}$ and $\mathcal{P}$ and therefore the NLS equation cannot predict FD in the spectral mean sense.  The Dysthe equation preserves $\mathcal{M}$, but does not necessarily preserve $\mathcal{P}$.  Since the sign in the change of $\mathcal{P}$ depends on the solution under consideration, the Dysthe equation predicts FD for some waves, frequency upshift for other waves, and constant spectral mean for other waves.  The dNLS equation does not preserve $\mathcal{M}$, nor does it preserve $\mathcal{P}$.  However, the dNLS equation preserves the spectral mean, $\omega_m$, so it cannot exhibit FD in the spectral mean sense.  This result is related to the fact that dissipation in the dNLS equation is frequency independent.  The vDysthe equation does not preserve $\mathcal{M}$ or $\mathcal{P}$.  The sign in the change of $\mathcal{P}$ is indefinite for the vDysthe equation just as it is for the Dysthe equation.  Therefore, the vDysthe equation can exhibit frequency downshift or upshift depending on the solution under consideration.  Finally, the dGT equation predicts FD in the spectral mean sense for all nontrivial wave trains.

In the remainder of this paper, we study the efficacy of these generalizations of the NLS equation at modeling the evolution of ocean swell as it travels across the Pacific Ocean.  In order to test the accuracy of these models, we focus on two questions: (i) how important are dissipation and decay? and (ii) how important are nonlinear effects?  These questions have been addressed using laboratory data, but to our knowledge, have not be addressed using ocean data. 

The remainder of the paper is outlined as follows.  Section \ref{OceanData} contains a description of the \cite{Snodgrass} ocean data and how we processed it.  Section \ref{Comparisons} contains the results of comparisons between the model predictions and the ocean data.  Finally, Section \ref{Conclusions} summarizes our observations and results.

\section{Ocean Data}
\label{OceanData}

\subsection{Description of Data}

\begin{figure}
  \centering
  \includegraphics[width=0.5\linewidth]{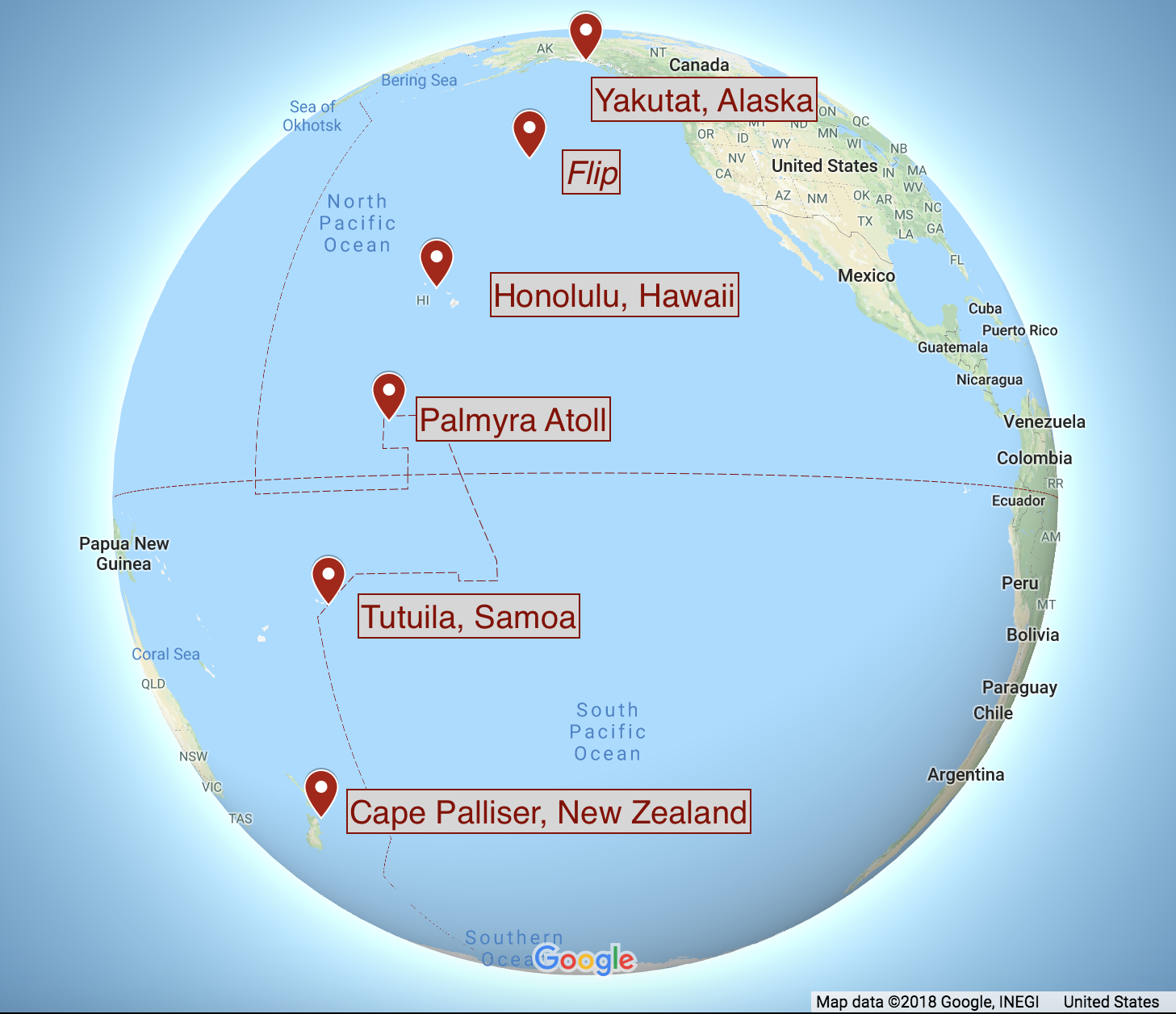}
  \caption{The six stations manned by~\cite{Snodgrass}. Each station collected power density spectra for swell propagating northward across the Pacific Ocean.  Map from~\cite{GoogleEarth}.}
  \label{fig:GreatCircle}
\end{figure}

During the southern hemisphere winter of 1963, a team of researchers led by Frank E.~Snodgrass and Walter Munk, both from the University of California Institute of Geophysics and Planetary Physics, collected wave data to measure the evolution of swell across the Pacific Ocean.  In order to track swell originating from storms in the southern hemisphere propagating northwards, the team operated six stations along a great circle, see Figure \ref{fig:GreatCircle}.  The locations included: Cape Palliser in New Zealand, Tutuila in American Samoa (14\textdegree22'S, 170\textdegree47'W), Palmyra Atoll (5\textdegree52'S, 162\textdegree7'W), Honolulu in Hawaii (21\textdegree14'S, 157\textdegree52'W), the vessel \textit{FLIP} in the north Pacific, and Yakutat in Alaska (59\textdegree30'N, 139\textdegree49'W).  However, the Cape Palliser and \textit{FLIP} locations did not produce data included in the current work. Due to dispersion, the ocean swell was recorded at any particular station for up to one week.  Three hours of time series pressure data were collected twice daily and converted to surface wave spectra.  Taking a ridge cut of these narrow-banded spectra at each station resulted in a composite spectrum removing the effects of dispersion.  \cite{Snodgrass} presented this data in the form of a power density spectrum, $C(f)$, with units of energy density (dB above 1 cm$^2$/mHz) per frequency (mHz) on a logarithmic scale.  They corrected the data to account for geometric spreading and island shadowing.  In addition, they determined the impact of effects such as instrument placement, refraction, oblateness of the earth, and wave-wave interactions.  They did not account for any forms of dissipation or decay.

\begin{figure}
    \centering
    \includegraphics[width=0.75\linewidth]{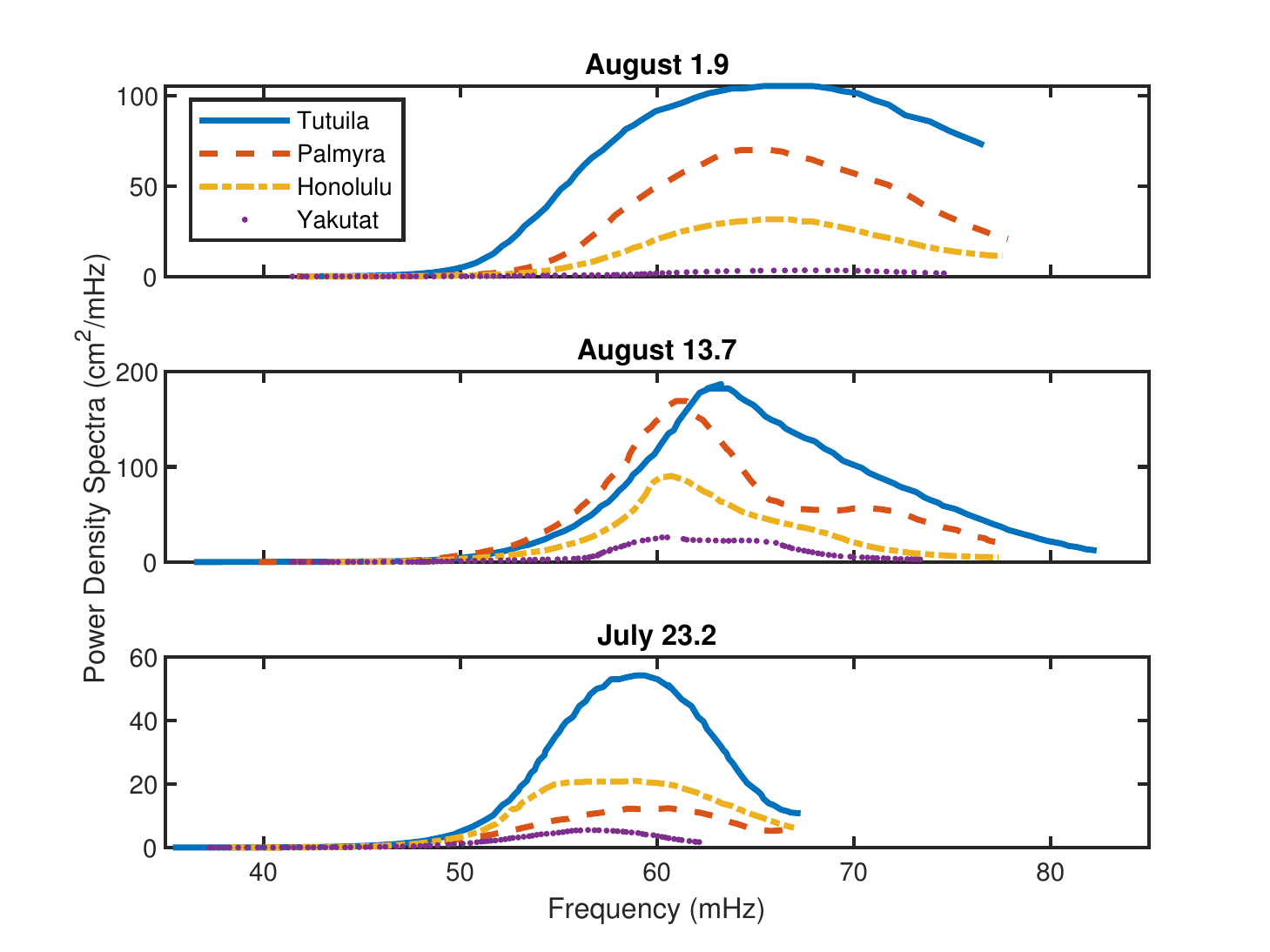}
    \caption{The power density spectra for the swells of August 1.9, August 13.7, and July 23.2.}
    \label{fig:spectra}
\end{figure}

Overall, swells from twelve storms were observed, but detailed ridge spectra were only provided for five swells.  In this study, we focus on the swells named August 1.9, August 13.7, and July 23.2.  For these three swell, data was only recorded by the gauges at Tutuila, Palmyra, Honolulu, and Yakutat.  The spectra corresponding to these swells are included in Figure~\ref{fig:spectra}.  The swells of August 13.7 and July 23.2 exhibit FD in the spectral mean sense, while only the swell of August 13.7 definitively exhibits FD in the spectral peak sense.  The swell of August 1.9 exhibits only temporary FD in the spectral mean sense, and its momentum, $\mathcal{P}$, increases as the waves propagate. 

A narrow-bandedness assumption is reasonable for these three swell. The second-to-last column of Table~\ref{tab:params} contains a measure of each swell’s narrow-bandedness, $\text{HWHM}\omega/\omega_0$, at the first gauge using half-width-half-maximum to determine $\text{HWHM}\omega$.  The values for August 13.7 and July 23.2 are both quite small, while the value for August 1.9 is reasonably small.  We do not consider the other two swells presented by~\cite{Snodgrass} because their spectra are not narrow banded or have multiple peaks, rendering them outside of the expected range of validity of the mathematical models considered herein. 

There are aspects of the swell data that limited our work. First, for the three swells considered in this study, data is only provided at four gauges. As we used the data at the first gauge to determine the initial conditions for our simulations, there were only three gauges to compare the simulation results against.  This limited our ability to make strong conclusions.  Second, data was only collected on a great circle so we only have a one-dimensional slice of the actual (two-dimensional) data.  Third, in the July 23.2 spectra, the energy at the third gauge is higher than the energy at the second gauge, which is evidence of uncertainty in the data.  Fourth, the data collection and processing techniques used by~\cite{Snodgrass} resulted in a loss of phase data, which is necessary to produce the physical surface displacement time series realizations required by the PDE models presented above.  Finally, the domain of frequencies present in each spectra varies across gauges.

\subsection{Data Processing}
\label{DataProcessingSection}

Given the power density spectra presented in \cite{Snodgrass} for a swell, to create initial conditions for our models, we created realizations of surface displacement time series at each gauge.  To do this, the data was digitized and interpolated to create a continuous spectrum since we did not have access to the original data.  Following the method presented in~\cite{Mobley}, we converted from decibels to units of energy density, cm$^2$/mHz, by taking $\Phi(f)=10^{C(f)/10}$.  Next, we discretized the continuous data into bands of width $\Delta f$, where $\Delta f=1/L$, with $L$ representing the collection period of three hours, and computed the magnitudes of the Fourier amplitudes, $|a|= 0.01\sqrt{\Phi(f) \Delta f/2},$ which have units of meters. The factor of $1/2$ comes from converting from one-sided to two-sided Hermitian spectra. We created a discretization grid around the spectral peak at the first gauge and ensured that each subsequent gauge maintained the same grid.  To compensate for the lack of phase data, each Fourier mode was assigned a random phase, preserving the magnitude of each amplitude. Assigning random phases is appropriate when the phase data is missing, see for example, \cite{Hol}.  We then generated a two-side Hermitian spectrum and took an inverse discrete Fourier transform to find a realization of the surface displacement time series.
 
\section{Model Comparisons}
\label{Comparisons}

\subsection{Computation of Parameters}
\begin{figure}
    \centering
    \includegraphics[width=0.9\linewidth]{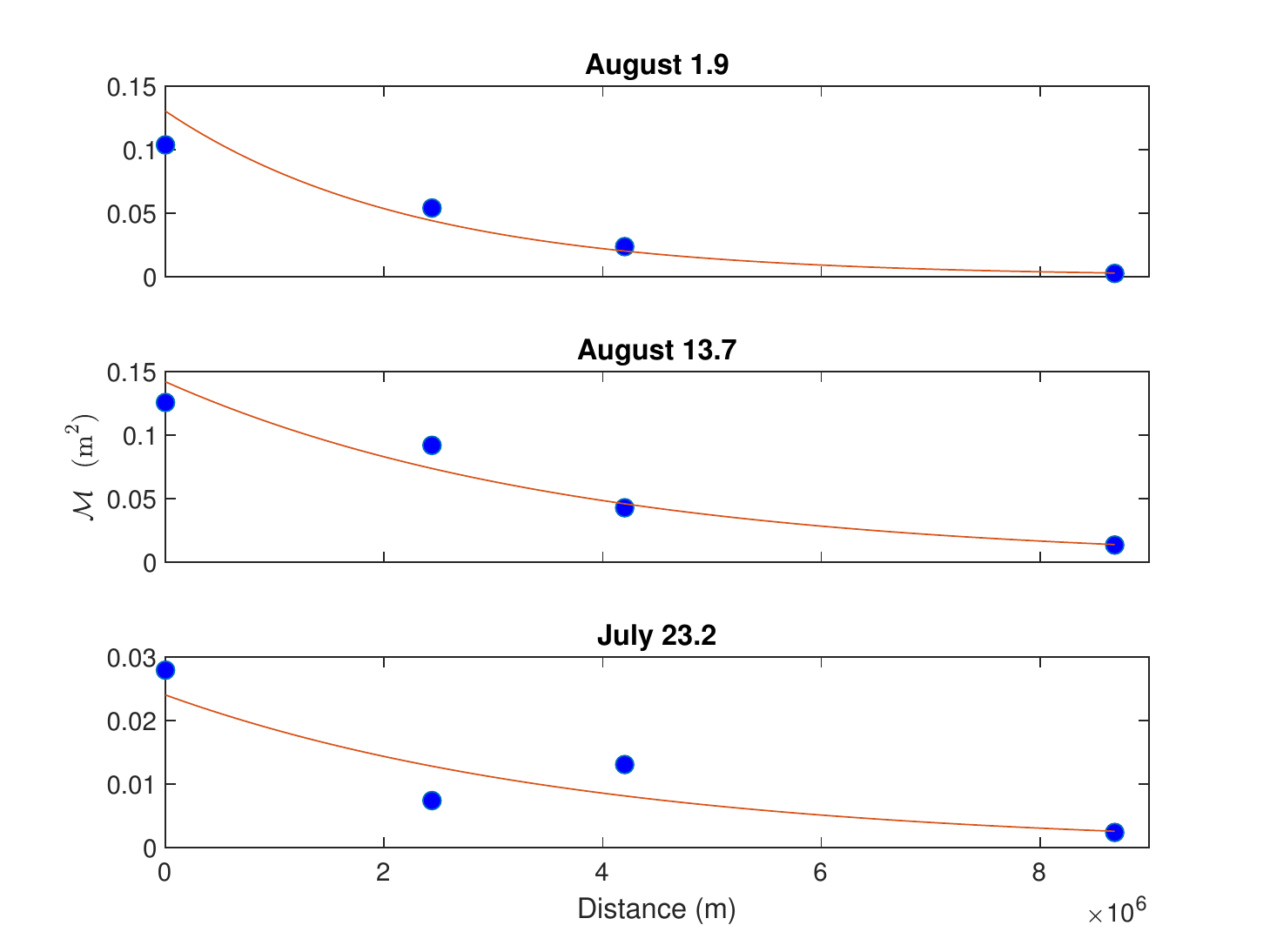}
    \caption{Plots of the dimensional mass, $\mathcal{M}$, versus the distance along the great circle.  The dots represent the physical measurements and the curve represents the best exponential fit.  From left to right, the dots refer to the stations in Tutuila, Palmyra, Honolulu, and Yakutat.}
    \label{fig:Mfit}
\end{figure}

There are two dimensionless parameters that appear in the models examined in this work.  The wave-steepness/nonlinearity parameter, $\epsilon$, is defined by $\epsilon=2a_0k_0$ where $a_0$ and $k_0$ are the amplitude and wavenumber of the carrier wave respectively.  The values of $\omega_0$, the frequency of the carrier wave, and $a_0$ were obtained directly from the spectrum at the first gauge.  The value of $k_0$ was determined using the deep-water linear dispersion relation, $\omega_0^2=gk_0$.  Table~\ref{tab:params} contains the values of these parameters for the three swell examined herein.  We note that our $\epsilon$ values are different than those in \cite{HendersonSegur} because we used a different definition for $\epsilon$.  This difference is irrelevant because the results we present below are independent of the value of $\epsilon$ due to an invariance of the PDEs.  The dissipation parameter, $\delta$, was determined by best-fitting an exponential function through the decay of $\mathcal{M}$ with respect to the dimensionless distance $\chi$, along the great circle.  Again, the parameter, $\delta$, accounts for {\emph{all}} dissipation and decay.  Figure~\ref{fig:Mfit} contains plots of dimensional $\mathcal{M}$ and the best exponential fit for each of the three swells.  The values of $\delta$ for each swell are included in Table~\ref{tab:params}.  For comparison, these $\delta$ values correspond to e-folding distances (the distance it takes the energy to change by a factor of $e$) of approximately 2,300~km (August 1.9, with a period of $T_0=$14.8~s), 3,700~km (August 13.7, with a period of $T_0=$15.8~s), and 3,900~km (July 23.2 with a period of $T_0=$16.9~s).  This shows that the e-folding distance increases with wave period, $T_0$.  Additionally, these values are consistent with the \cite{Collard} synthetic aperture radar measured e-folding distance of 3,300~km for a wave with a 15~s period. 

\begin{table}
\begin{center}
 \begin{tabular}{c| c c c c c c c c c} 
 Swell & $\omega_0$ (rad/s) & $T_0$ (s) & $k_0$ (m$^{-1}$)& $a_0$ (m) &  $\epsilon$  & $\mathcal{M}_0$ (m$^2$) & $\delta$   & $\text{HWHM}\omega/\omega_0$ & Downshift \\ 
 \hline
 August 1.9 & 4.25e-1 & 14.8 & 1.84e-2 & 2.21e-2 & 8.11e-4 & 1.04e-1       & 1.84e1 &  1.50e-1& NA\\ 
 August 13.7 & 3.97e-1 & 15.8 & 1.61e-2 & 2.94e-2 & 9.46e-4 & 1.26e-1   & 9.35e0 & 9.42e-2 & $\omega_m$, $\omega_p$\\
 July 23.2 & 3.72e-1 & 16.9 & 1.41e-2 &  1.59e-2 & 4.47e-4& 2.80e-2      & 4.58e1 & 8.32e-2 &  $\omega_m$\\
\end{tabular}
\caption {Parameters for each swell.  The parameters $\omega_0$, $T_0$, $k_0$, and $a_0$ represent the angular frequency, temporal period, spatial wavenumber, and the amplitude of the carrier wave, respectively.  The parameter $\epsilon$ is the dimensionless nonlinearity parameter, $\mathcal{M}_0$ is the value of $\mathcal{M}$ at the first gauge (Tutuila), and $\delta$ is the dimensionless dissipation parameter.  The second-to-last column shows a measure of the narrow bandedness of the spectra.  The final column shows whether FD in the spectral mean, $\omega_m$, or spectral peak, $\omega_p$, sense occurred in each swell as it propagated northwards.}
\label{tab:params} 
\end{center}
\end{table}

\subsection{Simulation Methods}
All model PDEs were solved numerically in dimensionless form by assuming periodic boundary conditions in $\xi$ and using the sixth-order operator splitting algorithm developed by \cite{Yoshida} in $\chi$ in Python and Fortran.  The linear parts of the PDEs were solved exactly in Fourier space using the fast Fourier transform (FFT).  The nonlinear parts of the PDEs were either solved exactly (NLS, dNLS) or using fourth-order Runge-Kutta (Dysthe, vDysthe, dGT) in physical space.  We used dimensionless initial conditions of the form
\begin{equation}
    u(\xi,\chi=0)=\frac{k_0}{\epsilon}\sum_j 0.01\sqrt{\frac{\Phi(f_j) \Delta f}{2}}~\exp(i\theta_j) \exp\left(\frac{2\pi ij}{\epsilon\omega_0L}\xi\right),
\end{equation}
where $\Phi(f)$ is the power density spectrum at the first gauge (Tutuila), $j=0$ corresponds to the spectral peak at the first gauge, $\theta_j\in [0,2\pi]$ is a random, uniformly distributed value for each frequency, and $L$ is the data collection period of three hours.  The distances between Tutuila and the remaining gauges were approximated to be 2.4e6~m (Palmyra), 4.2e6~m (Honolulu), and 8.7e6~m (Yakutat).  Thus, the PDEs were solved on the (dimensionless) interval $\chi\in[0,8.7*10^6\epsilon^2k_0]$.  As a check of simulation accuracy, the evolution of the quantities $\mathcal{M}$ and $\mathcal{P}$ was compared against theoretical predictions and was found to be consistent, indicating that the implemented numerical methods correctly solved each PDE.  However, note that all further reported values of $\mathcal{M}$, $\mathcal{P}$, and $\omega_m$ at each gauge were computed using only frequencies present at that gauge. See Section \ref{SectionResults} for a discussion of the implications of this restriction.

The simulation results were re-dimensionalized and compared with the ocean swell measurements at each gauge using the error norm 
\begin{equation}
    \mathcal{E}=\sum_{n=2}^{4} \sum_{j=-J_n}^{J_n}  \frac{1}{3\mathcal{M}}_n\bigg| \big|\hat{B}_{n}^{\text{sim}}(j)\big|-\big|\hat{B}^{\text{data}}_n(j)\big|\bigg|^2,
    \label{eqn:error}
\end{equation}
where $n$ represents the gauge number, where $J_n$ is the number of nonzero Fourier modes at gauge $n$, $\mathcal{M}_n$ is the value of $\mathcal{M}$ at the $n^{th}$ gauge, and $\hat{B}_n(j)$ is the $j^{th}$ nonzero Fourier amplitude at the $n^{th}$ gauge from the numerical simulation (sim) or the ocean swell data (data). This process was repeated 200 times with different random phases for each swell. The mean of the results is reported to account for random effects. Additionally, to compare nonlinear and linear theories, solutions to both the full (nonlinear) PDEs and their linearizations were computed.  Note that phase does not affect the linear results, so only one random phase simulation was computed for each linearized PDE for each swell.  Also note that since $\mathcal{M}$ typically decreases as the waves travel north, the error contribution from the last gauge (Yakutat) will be more pronounced than the error contributions from the other gauges.

\subsection{Results}
\label{SectionResults}
\begin{figure}
    \centering
    \includegraphics[width=\linewidth]{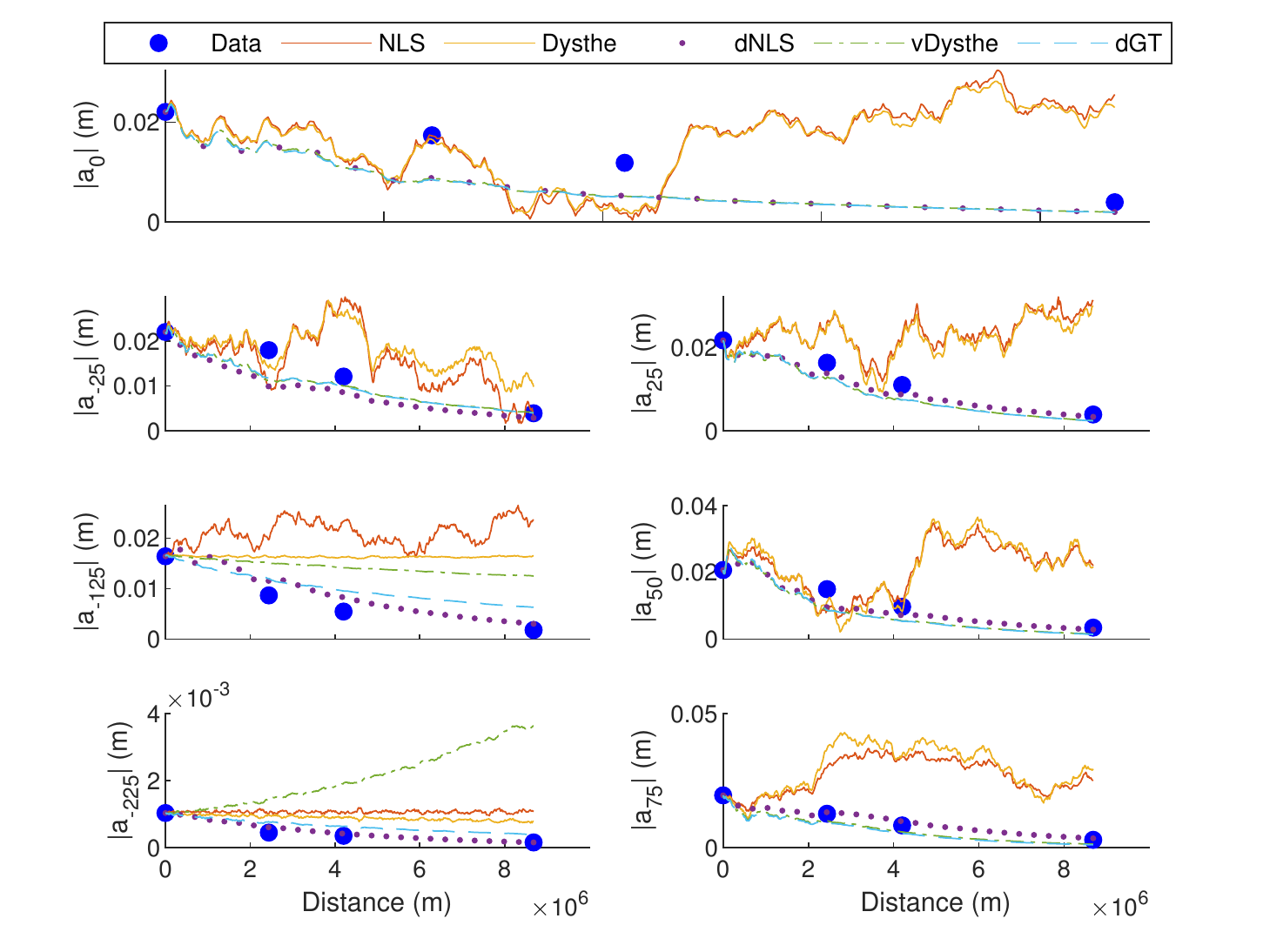}
\caption{Plots of seven Fourier amplitudes versus distanced traveled comparing the nonlinear PDE predictions from a single random-phase simulation (curves) with the ocean data (big blue dots) for the August 1.9 swell.  The top plot is of the carrier wave amplitude.  The left column contains plots of three lower sideband amplitudes and the right column contains plots of three upper sideband amplitudes.}
    \label{fig:result}
\end{figure}

\begin{figure}
    \centering
    \includegraphics[width=\linewidth]{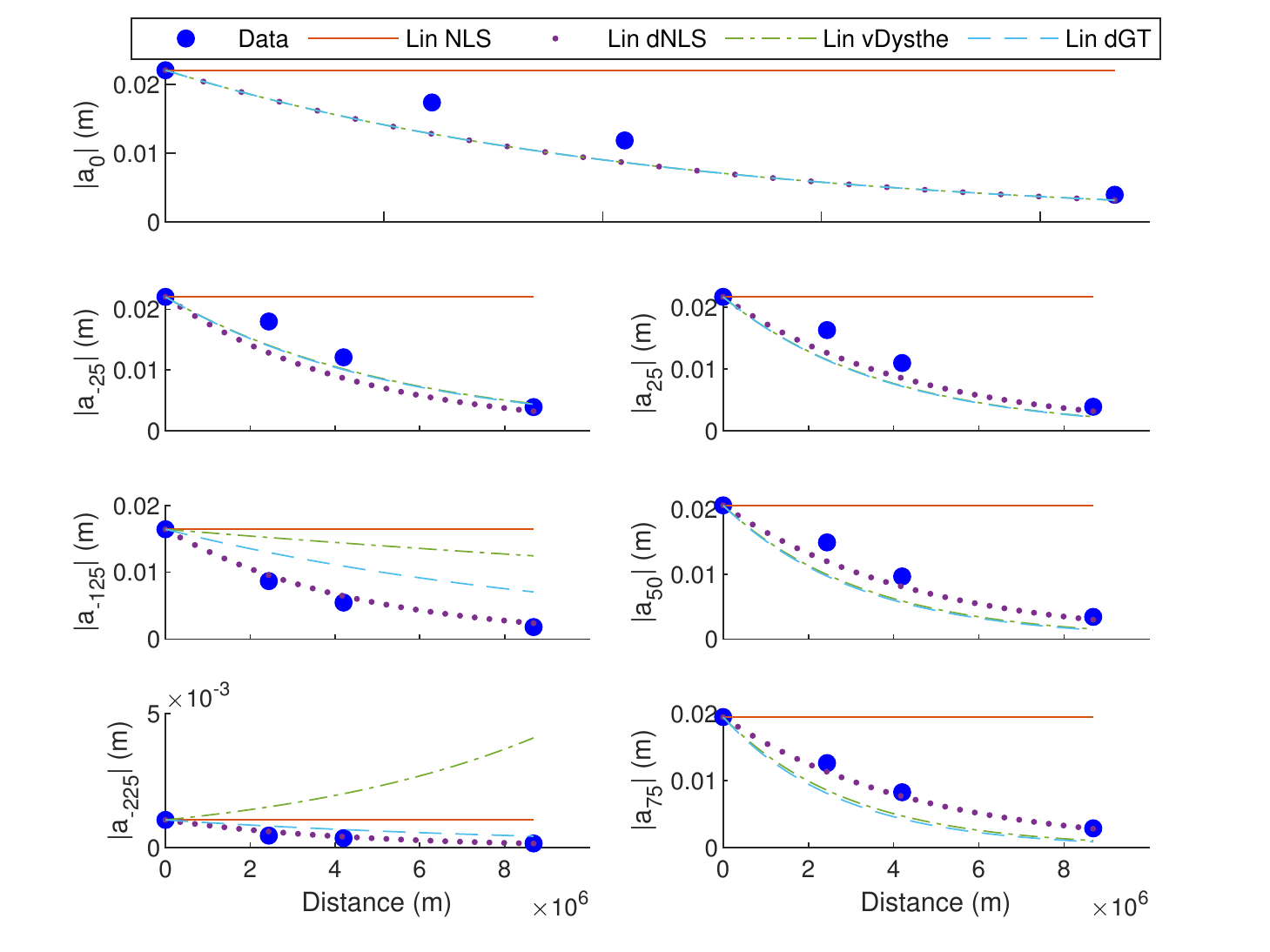}
\caption{Plots of seven Fourier amplitudes versus distanced traveled comparing the {\emph{linearized}} PDE predictions from a single random-phase simulation (curves) with the ocean data (big blue dots) for the August 1.9 swell.  The top plot is of the carrier wave amplitude.  The left column contains plots of three lower sideband amplitudes and the right column contains plots of three upper sideband amplitudes.}
    \label{fig:linearresult}
\end{figure}

Figure~\ref{fig:result} shows plots that compare the August 1.9 ocean swell data with the numerical predictions from the {\emph{nonlinear}} PDEs for the amplitudes of the carrier wave and six sidebands.  Figure~\ref{fig:linearresult} shows similar comparisons for the {\emph{linearized}} PDEs.  Since the linearizations of the NLS and Dysthe equations are the same, only the linearized NLS results are explicitly shown.  The sidebands shown represent a broad range of the swell's spectrum and demonstrate the nonphysical exponential growth in the far lower sidebands predicted by the vDysthe and linearized vDysthe equations.  A number of observations can be made from these figures.  First, the dissipative models (dNLS, vDysthe, dGT) are generally much more accurate than the conservative models (NLS, Dysthe).  Second, the NLS and Dysthe predictions are quite similar.  Third, the ``curves'' corresponding to the linearized NLS/Dysthe equation in Figure \ref{fig:linearresult} are horizontal lines.  This shows that energy is not transferred between modes in these linearized models.  Fourth, there are strong qualitative similarities between the predictions from a nonlinear model and its linearization.  The plots for the August 13.7 and July 23.2 swells (not shown) are qualitatively similar.

\begin{figure}
    \centering
    \includegraphics[width=\linewidth]{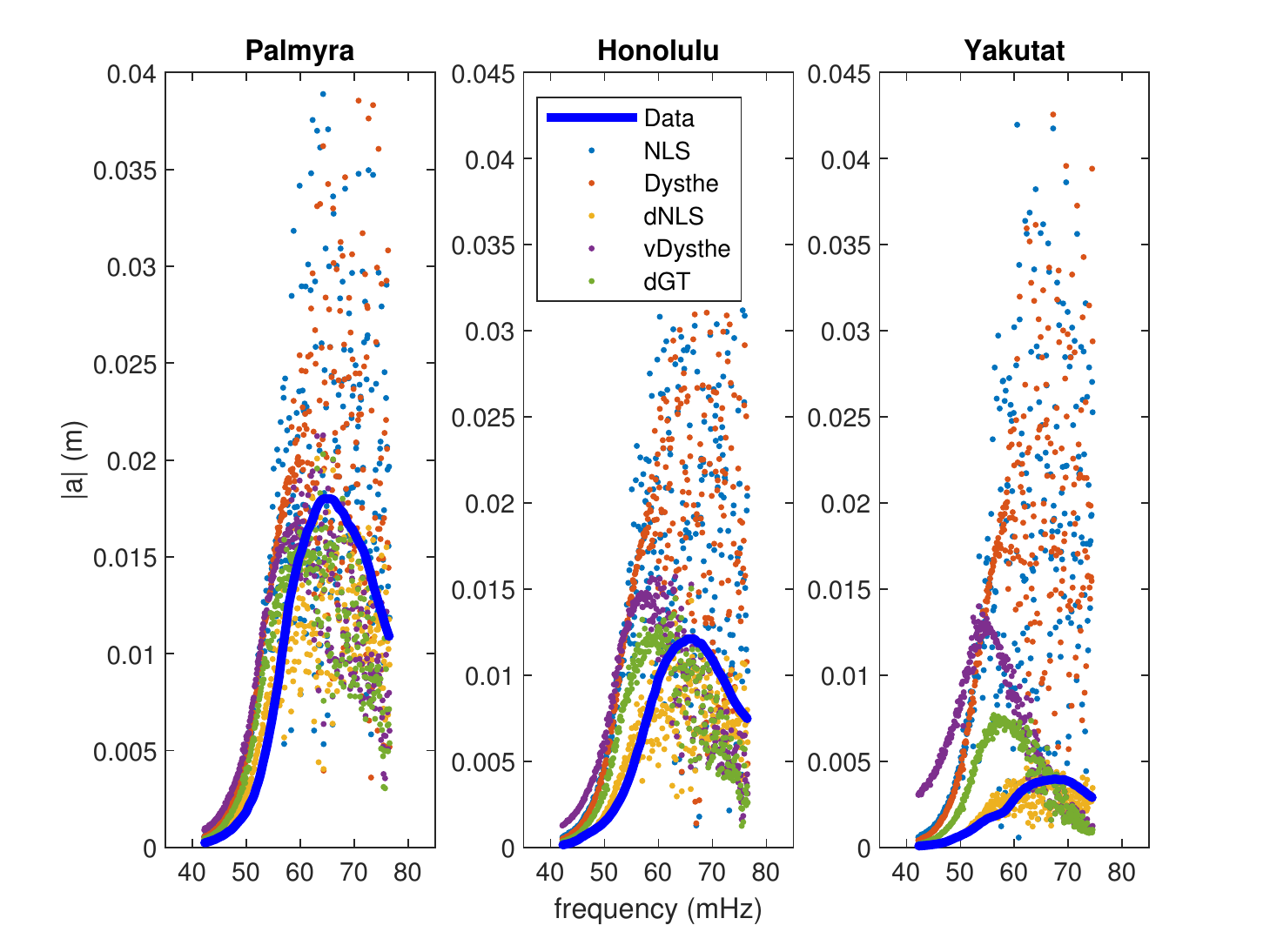}
\caption{Plots of the amplitude spectra comparing the PDE predictions (colored dots) from a single random-phase simulation with the August 1.9 swell data (blue curves) at the three downstream gauge locations.}
    \label{fig:spectraresult}
\end{figure}

Figure~\ref{fig:spectraresult} includes plots that compare the complete August 1.9 ocean data amplitude spectra with the amplitude spectra from one random-phase simulation of the nonlinear PDEs.  The comparisons are only shown at the three ``downstream'' gauge locations because the first gauge (Tutuila) was used to define the initial conditions for the simulations and therefore the ocean data and numerical predictions match exactly there.  \cite{Snodgrass} did not report the power density spectra using the same frequency interval for all gauges and all swells.  In order to be consistent with this fact, in Figure~\ref{fig:spectraresult} and all following figures, we only report model predictions from the frequency intervals for which we have ocean data (even though our numerical simulations contained many more modes).  Figure~\ref{fig:spectraresult} shows that there are significant variations in the amplitudes of neighboring frequencies for a given random-phase simulation.  These variations depend on the random phases used in the initial conditions.  Regardless, similarly to the plots in Figures~\ref{fig:result} and~\ref{fig:linearresult}, these plots show that the dissipative models are much more accurate than the conservative models.

\begin{figure}
    \centering
    \includegraphics[width=\linewidth]{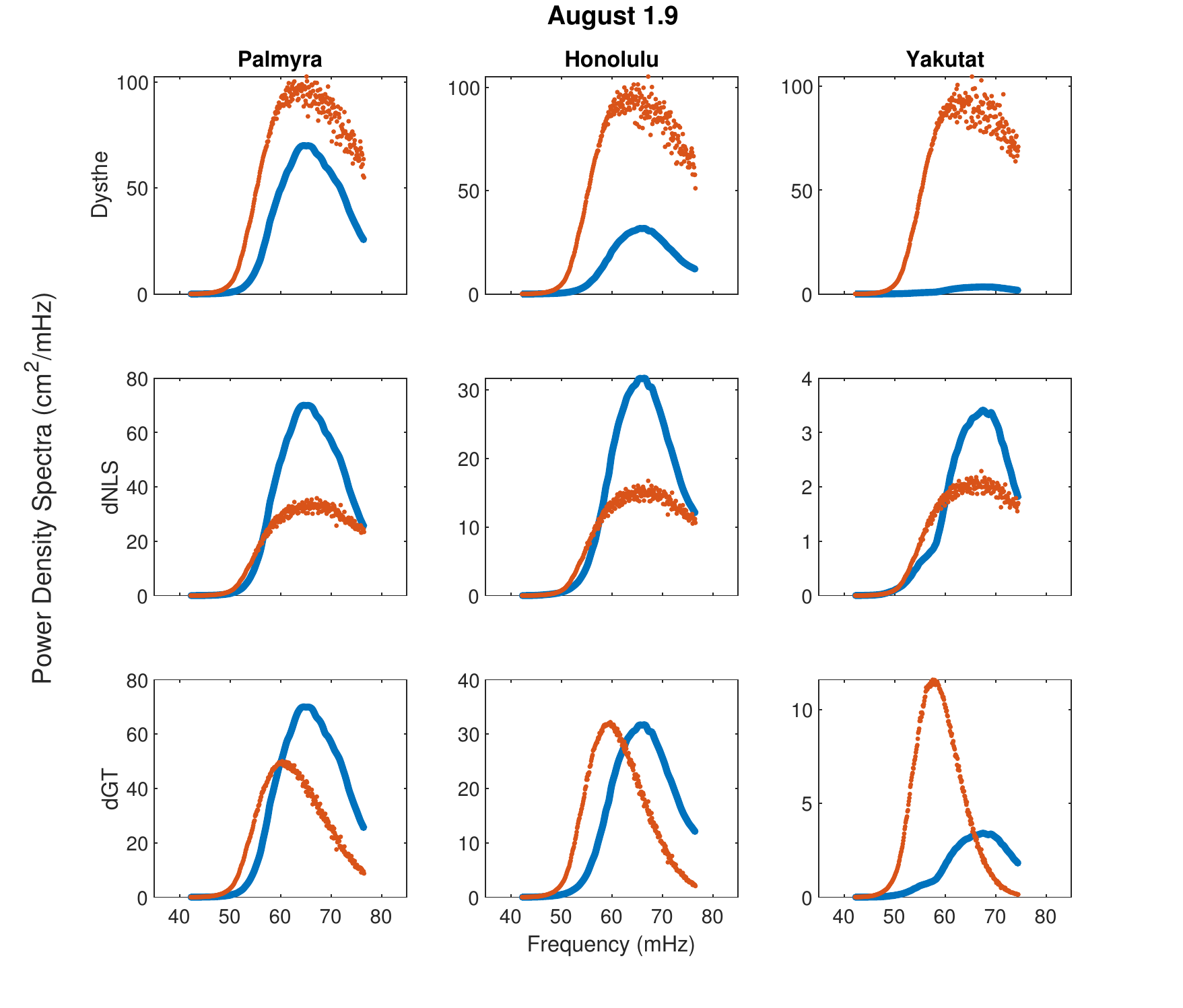}
\caption{Plots of the power density spectra for the August 1.9 swell comparing the ensemble-averaged PDE predictions (orange dots) with the ocean swell data (thick blue curves).  The rows contain the Dysthe, dNLS, and dGT predictions.  The columns contain the results at the Palmyra, Honolulu, and Yakutat gauges.}
    \label{fig:spectraresultAug1}
\end{figure}

\begin{figure}
    \centering
    \includegraphics[width=\linewidth]{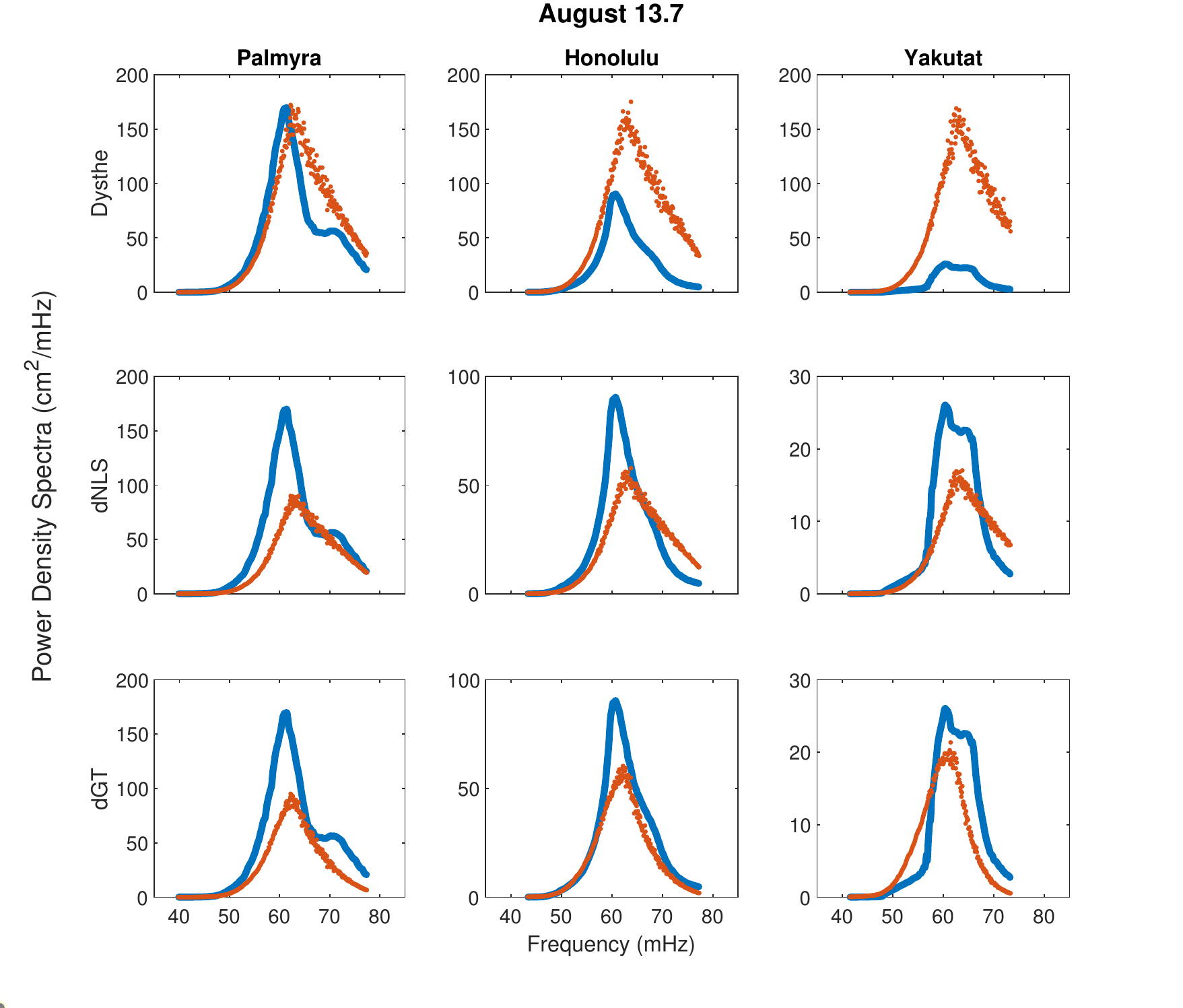}
    \caption{Plots of the power density spectra for the August 13.7 swell comparing the ensemble-averaged PDE predictions (orange dots) with the ocean swell data (thick blue curves).  The rows contain the Dysthe, dNLS, and dGT predictions.  The columns contain the results at the Palmyra, Honolulu, and Yakutat gauges.}
    \label{fig:spectraresultAug2}
\end{figure}

\begin{figure}
    \centering
    \includegraphics[width=\linewidth]{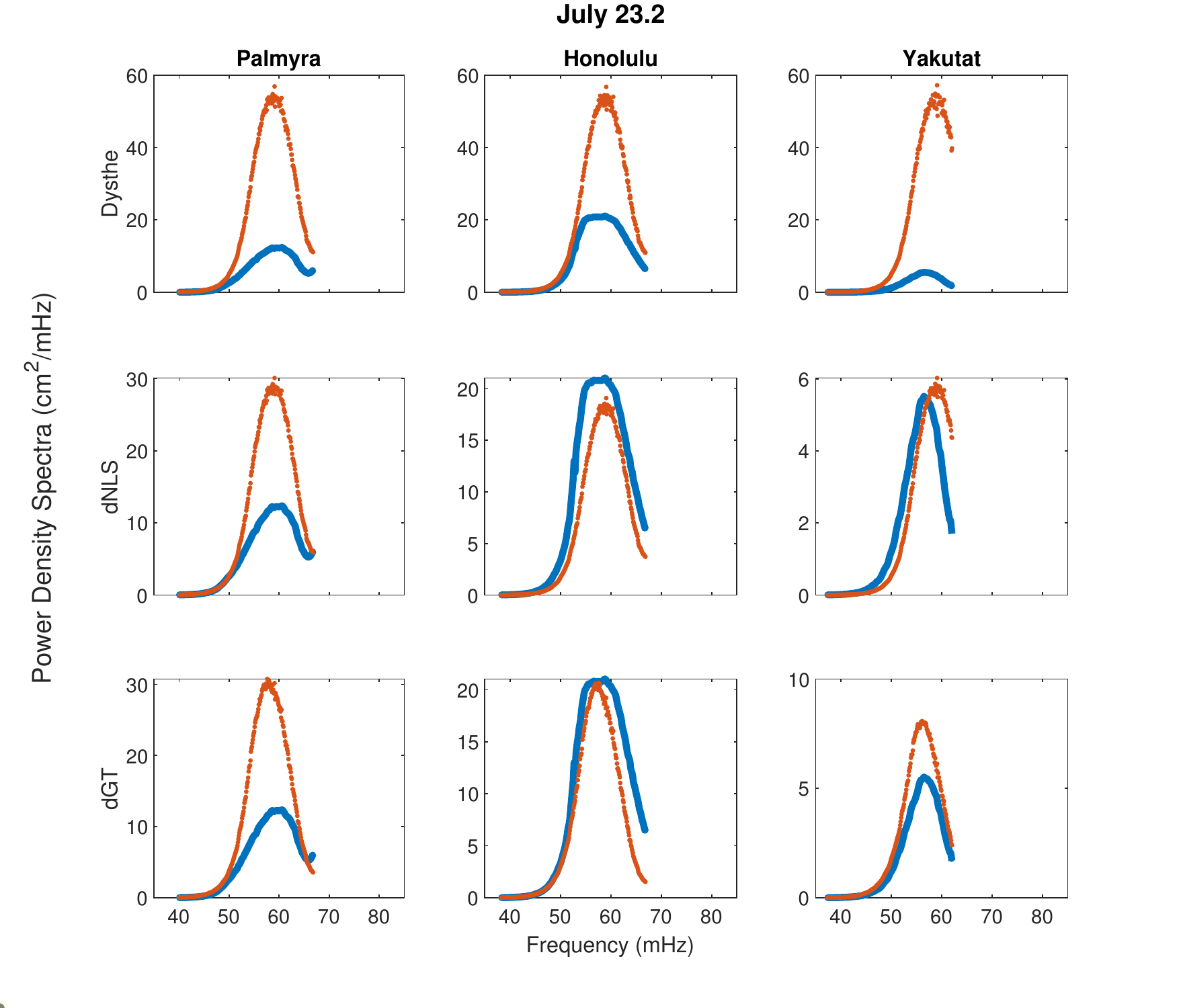}
    \caption{Plots of the power density spectra for the July 23.2 swell comparing the ensemble-averaged PDE predictions (orange dots) with the ocean swell data (thick blue curves).  The rows contain the Dysthe, dNLS, and dGT predictions.  The columns contain the results at the Palmyra, Honolulu, and Yakutat gauges.}
    \label{fig:spectraresultJuly}
\end{figure}

We emphasize that Figure~\ref{fig:spectraresult} contains results from {\emph{a single}} random-phase simulation.  Each different random-phase simulation gives different results.  Since there are variations between each random-phase simulation, we averaged the simulation values over an ensemble of 200 random-phase simulations.  Figures \ref{fig:spectraresultAug1}-\ref{fig:spectraresultJuly} show comparisons of the ensemble-averaged power density spectra predictions from the PDEs and the ocean for the swell of August 1.9, August 13.7, and July 23.2, respectively.  For conciseness, only the results from the Dysthe, dNLS, and dGT equations are included.  The results from the NLS equation are roughly similar to the results of the Dysthe equation and the results from the vDysthe equation are roughly similar to those of the dGT equation.  These plots show that the NLS and Dysthe equations greatly overpredict the ocean measurements and that these predictions worsen as the waves travel north.  This is because these models do not include any dissipative or decay effects.  The accuracy of the dissipative models varies from swell to swell.  For the August 1.9 swell, dNLS consistently underpredicts the ocean measurements and dGT switches from underprediction to overprediction as the waves travel north.  Additionally, the dGT equation predicts a decrease in the spectral peak and mean that is not observed in the ocean data.  For the August 13.7 swell, both the dNLS and dGT equations underpredict the ocean measurements, though the dGT predictions are more accurate.  For the July 23.2 swell, both dissipative models overpredict the Palmyra measurements, but do reasonably well at the Honolulu and Yakutat gauges.  We note that if we interchange the Palmyra and Honolulu ocean data sets, then the agreement between the dNLS and dGT predictions and the ocean data increases significantly.  (Recall that the the $\mathcal{M}$ measurements make it appear like the Palmyra and Honolulu data sets have been switched.  See the bottom plot in Figure \ref{fig:Mfit}.)  In this swell, the dGT equation accurately predicts the decrease in the spectral peak, while the dNLS equation does not.

\begin{figure}
    \centering
    \includegraphics[width=\linewidth]{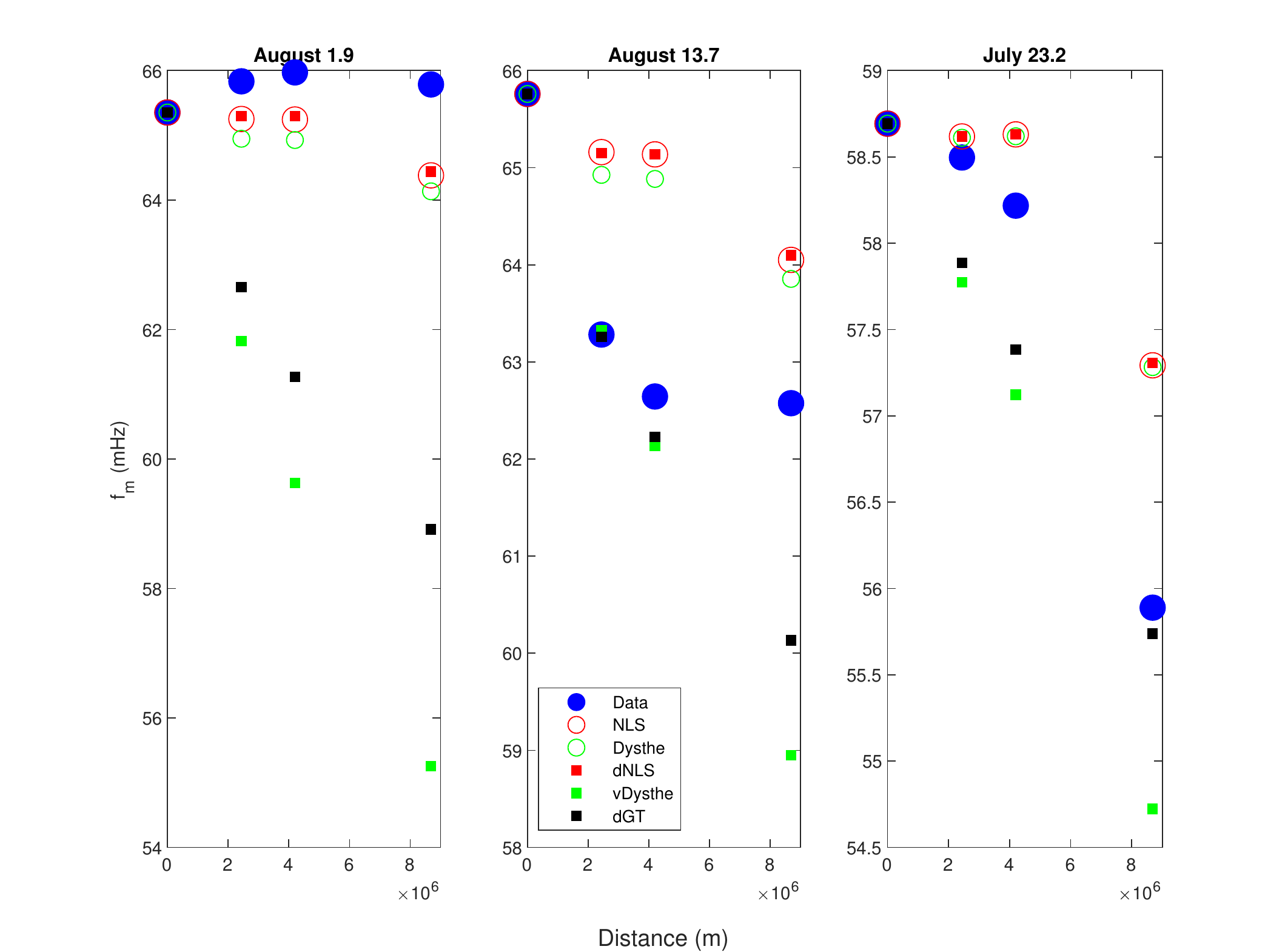}
\caption{Plots comparing the evolution of the mean frequency, $f_m$, for the ocean data and the ensemble-averaged model results for all three swells.}
    \label{fig:spectraMeanresult}
\end{figure}

Figure~\ref{fig:spectraMeanresult} compares the evolution of the ensemble-averaged mean frequencies, $f_m$, with the mean frequencies from the ocean swell data.  Recall that $f_m=\omega_m/(2\pi)$.  It is important to note that only the frequencies in the range reported by~\cite{Snodgrass} at each gauge were used.  This causes a decrease in the mean frequencies of the NLS and dNLS predictions that would not appear if all simulation frequencies were used.  The NLS, Dysthe, and dNLS equations most accurately model the evolution of $f_m$ for the (nondownshifting) August 1.9 swell.  For the August 13.7 and July 23.2 swells, the NLS, Dysthe, and dNLS equations underpredict the downshift while the vDysthe and dGT equations overpredict the downshift.  

Quantitative comparisons between the ocean swell data and simulations of the nonlinear PDEs using the error norm given in equation (\ref{eqn:error}) are reported in Table~\ref{tab:200runs}.  Quantitative comparisons between the ocean data and simulations of the linearized PDEs are reported in Table~\ref{tab:linear}.  For all three swells, the dissipative models performed between one and two orders of magnitude better than the conservative models.  This result is predictable because the spectra shown in Figure~\ref{fig:spectra} show that the swells generally lose energy as they traveled northwards.  These results further confirm that including dissipation is necessary to accurately model the evolution of swell as it travels across the Pacific Ocean.

Amongst the nonlinear PDEs, dNLS produced the smallest error for the August 1.9 swell; vDysthe produced the smallest error for the August 13.7 swell; and vDysthe produced the smallest error for the July 23.2 swell, though dGT produced a very similar result.  The differences between the linear and nonlinear results were small.  Out of all models considered, the linearized dNLS equation performed best for the swell of August 1.9, the vDysthe equation performed best for the swells of August 13.7 and July 23.2.  These results suggest that including nonlinear effects is not necessary to accurately model the evolution of swell across the Pacific.  However, because the differences between the linear and nonlinear predictions occur over (relatively) short distances, we hypothesize that the linear models sometimes appear more effective than the nonlinear models due the very low spatial resolution in the data. Ideally, we would compare our models against higher resolution data to determine the importance of nonlinearity.

\begin{table}
    \centering
    \begin{tabular}{c|c|c|c}
       Model    & August 1.9 & August 13.7 & July 23.2 \\
       \hline
       NLS      & 8.29       & 1.19      & 1.48\\
       Dysthe   & 8.76       & 1.19      & 1.49\\
       dNLS     & 0.0559     & 0.310     & 0.197\\
       vDysthe  & 2.02       & 0.149     & 0.145\\
       dGT      & 0.460      & 0.182     & 0.146
    \end{tabular}
    \caption{Averaged error results for ensembles of 200 random-phase simulations of the nonlinear PDEs using the error norm defined in equation (\ref{eqn:error}). }
    \label{tab:200runs}
\end{table}

\begin{table}
    \centering
    \begin{tabular}{c|c|c|c}
       Model                         & August 1.9   & August 13.7      & July 23.2 \\
       \hline
        Linearized NLS/Dysthe        & 10.1      & 1.47           & 1.18\\
        Linearized dNLS              & 0.0446    & 0.319          & 0.269\\
        Linearized vDysthe           & 2.04      & 0.159          & 0.168\\
        Linearized dGT               & 0.453     & 0.192          & 0.148        
    \end{tabular}
    \caption{Error results for simulations of the linearized PDEs using the error norm defined in equation (\ref{eqn:error}).  Note that the linearized versions of the NLS and Dysthe equations are the same.}
    \label{tab:linear}
\end{table}

\medskip

\noindent \textbf{Other observations:}
\begin{itemize}
    \item{The swell of August 13.7 had the most energy, while the swell of July 23.2 had the least. The fact that the linear models performed similarly to the nonlinear models in both of these cases suggests either that a swell needs to have even more energy than the August 13.7 swell for nonlinearity to be important or that there is not a simple relationship between energy and the importance of nonlinearity.}
    \item{All three of the swells had comparable carrier wave frequencies, and the slight variations do not appear to strongly affect the simulation predictions.}
    \item{All three of the swells were relatively narrow banded and the degree of narrow-bandedness does not appear to strongly affect the accuracy of the simulation predictions.}
    \item{None of the models accurately modeled the evolution of the spectral mean.  When downshift in the spectral mean sense occurred (as in the August 13.7 and July 23.2 swells), the conservative models underestimated the downshift and the dissipative models overestimated it.  When downshift did not occur (as in the August 1.9 swell), the conservative models most accurately predicted the evolution of the spectral mean.}
    \item{Because the domain of frequencies in the~\cite{Snodgrass} data changes for each gauge and each swell, only frequencies present at each gauge were used to calculate the simulation spectral mean.  This leads to behaviors different than theoretically predicted if all frequencies were included for the NLS and dNLS equations.}
    \item{Theoretically, swell exhibiting FD in the spectral mean sense should be best predicted by the vDysthe and dGT equations, which can both predict this phenomena.  However, we found that for the three swells examined herein, they were not consistently better than the dNLS predition.  We do not know why some swell downshift and others do not.}
    \item{The vDysthe and linearized vDysthe equations predict nonphysical exponential growth in lower sidebands that are further than $5/(\epsilon k_0)$ away from the carrier wave, see the lower left subplot in Figures~\ref{fig:result} and~\ref{fig:linearresult}.  The amplitudes of these modes was small enough that their exponential growth did not significantly increase the value of the error, $\mathcal{E}$}.
    \item{According to \cite{Snodgrass}, the swell of July 23.2 had more energy in Honolulu than in Palmyra.  This means that energy did not decay monotonically as the swell propagated northwards.  Switching the order of the data from these two gauges (so that the energy decays monotonically) does not have a large impact on the qualitative results.  However, switching the order causes the accuracy of the vDysthe and dGT predictions to increase significantly.}
    \item{We attempted to test the accuracy of the \cite{IslasSchober} model.  However, we found that the optimal value of their free parameter $\beta$ was negative, violating that model's assumptions. Thus, it is not a good model for these three swells.}    
\end{itemize}

\section{Conclusions}
\label{Conclusions}

We compared the ocean swell data collected by \cite{Snodgrass} with predictions from the nonlinear Schr{\"o}dinger, Dysthe, dissipative nonlinear Schr{\"o}dinger, viscous Dysthe, and dissipative Gramstad-Trulsen equations.  As only amplitude data was provided, we made the random phase assumption, ran 200 simulations with different random phases, and averaged the error between the ocean measurements and PDE predictions.  We found that the dissipative models (dNLS, vDysthe, dGT) performed orders of magnitude better than the conservative models (NLS, Dysthe), suggesting that dissipation is a physically important effect for swell propagating across the Pacific Ocean.  Additionally, for swells exhibiting frequency downshift in the spectral mean sense, models that can predict this behavior (vDysthe, dGT) performed the best.  The dissipative models, which are based upon a simple dissipation ansatz that combines all dissipation and decay into one parameter, provided good predictions for the swells as they propagated across the ocean.  We also found that the linear models performed slightly better than the nonlinear models.  This suggests that (dissipative) linear models may be sufficient for modeling some aspects of the evolution of swell across the Pacific.
 
\section*{Acknowledgements}
We thank Harvey Segur and Diane Henderson for helpful conversations and for providing us with the digitization of the \cite{Snodgrass} data.  We thank the referees for help comments and suggestions.  Additional thanks to Curtis Mobley, Bernard Deconinck, Christopher Ross, Hannah Potgieter, and Salvatore Calatola-Young for useful discussions.  Datasets, including Python codes, for this research are available through Harvard Dataverse at https://doi.org/10.7910/DVN/XWJENO.  This research was supported by the National Science Foundation under grant number DMS-1716120.

\end{document}